\documentclass[twocolumn,secnumarabic,amssymb, nobibnotes, aps, prl]{revtex4-1}

\usepackage{graphicx}
\usepackage{amsmath}
\usepackage{enumitem}
\usepackage{siunitx}
\usepackage{bm}
\usepackage[colorlinks=true,citecolor=blue,linkcolor=blue]{hyperref}
\newlist{eqlist}{enumerate*}{1}
\makeatletter
\setlist[eqlist]{itemjoin=\quad,mode=unboxed,label=(\roman*),
  ref=\theequation(\roman*),before={\let\label\ltx@label}}
\makeatother

\begin{document}

\title{Lift induced by slip inhomogeneities in lubricated contacts}%

\author{Aidan Rinehart$^1$}%
\author{U\v{g}is L\={a}cis$^1$}
\author{Thomas Salez$^{2,3}$}
\author{Shervin Bagheri$^1$}

\email[corresponding author: ]{shervin@mech.kth.se}
\affiliation{$^1$Dept.~Engineering Mechanics, KTH Royal Institute of Technology, Stockholm SE-10044}
\affiliation{$^2$Univ.~Bordeaux, CNRS, LOMA, UMR 5798, F-33405, Talence, France}
\affiliation{$^3$Global Station for Soft Matter, Gi-CoRE, Hokkaido University, Sapporo, Hokkaido 060-0808, Japan}

\date{November 2019}%
\begin{abstract}
Lubrication forces depend to a high degree on elasticity, texture, charge, chemistry, and temperature of the interacting surfaces. Therefore, by appropriately designing surface properties, we may tailor lubrication forces to  reduce friction, adhesion and wear between sliding surfaces or control repulsion, assembly, and collision of interacting particles. Here, we show that variations of slippage on one of the contacting surfaces induce a lift force. 
We demonstrate the consequences  of this force on the mobility of a cylinder traveling near a wall and show the emergence of particle oscillation and migration that would not otherwise occur in the Stokes flow regime.
Our study has implications for understanding how inhomogeneous biological interfaces interact with their environment; it also reveals a new method of patterning surfaces for controlling the motion of nearby particles. 

\end{abstract}

\maketitle

In many physical processes, the flow of small particles such as cells, colloids, bubbles, grains, and fibers occurs near soft, porous and rough walls. The induced lubrication forces \cite{reynolds} on these particles depend on the elasticity, texture, and chemistry of the nearby wall. These forces may dominate over both bulk (e.g.~Stokes drag) and surface (e.g.~van der Waals and electrostatic) forces, and  therefore determine single-particle motion and collective behavior.

The simplest configuration to characterize hydrodynamic particle-wall forces is that of an infinitely long circular cylinder traveling parallel to a rigid flat wall.  At low Reynolds numbers, a rigid cylinder will experience zero wall-normal (lift) force and therefore move at a constant distance from the wall \cite{jeffery}. This  is due to the time-reversal symmetry of the Stokes equations. However, if one of the interacting surfaces is soft, the moving particle will be repelled from the wall \cite{sekimoto1993softlubrication} as a result of the broken symmetry of the fluid pressure in the thin gap \cite{cantat_PRL_1999, Abkarian_PRL, beaucourt2004elasticlift, skotheim2004soft, urzay2007elastohydrodynamiclift, snoeijer2013lubricated_contact}.
This elastohydrodynamic lift mechanism increases the gap thickness and reduces wear and friction between the siding surfaces \cite{Saintyves, rallabandi2018elasticmembrane, Vialar}. 
It  underlies exotic particle trajectories such as oscillations, Magnus-like effect, stick-slip motion, and spinning \cite{salez_mahadevan_2015, Rallabandi_rotatingelastic}. 
Soft lubrication  also underpins surface rheology   \cite{Steinberger_bubble_elastic2008,Leroy_hydro_elasticfilm, wang2017elastic}  that is used to characterize the viscoelasticity of complex surfaces.

Besides softness, another ubiquitous feature of surfaces in biology and technology is surface inhomogeneities. 
For example, the surface of a Janus particle is divided into two halves with different chemistry (hydrophobic/hydrophilic) or texture (rough/smooth). This provides the particle with unique capabilities including self-assembly into complex structures \cite{Jing_janus_selfassembly} and self-propulsion \cite{Golestanian2005}. More generally,
interfaces in living tissues (cell walls, blood vessels, cartilage, epithelia) vary in chemical and mechanical composition due to inhomogeneous distribution of cells and proteins. In technological applications, inhomogeneous surfaces arise as a consequence of manufacturing imperfections and wear, but also from surface patterning to  control liquid transport \cite{WexlerSM} or heat transfer \cite{phase-change-2014}.  Despite this ubiquity, there has been no investigation of the full set of lubrication forces arising from particle-wall interaction when the properties of one of the contacting surfaces vary. 

\begin{figure}[t]
 \centering
        \includegraphics[width=1.02\columnwidth]{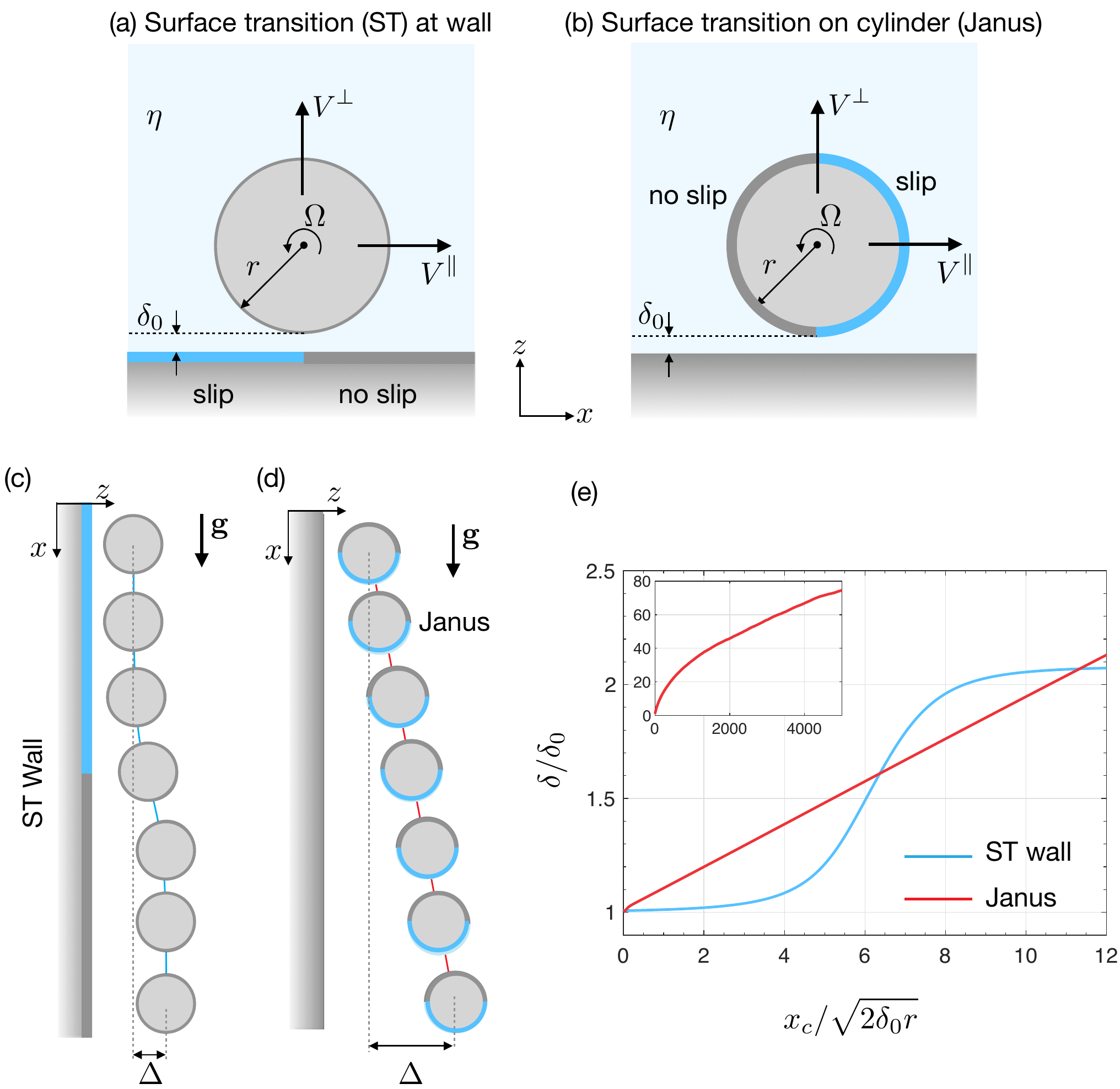}
 \caption[]{
Two lubrication models.  (a) 2D solid cylinder moving near a flat wall with a transition from a slip (blue) to a no slip  (grey). (b) 2D cylinder half-coated with a slip region and moving near a wall. (c),(d) Sketches of  cylinder trajectories for the two systems placed in a gravity field with $\delta_0 = 0.05r$, $\ell = 1.79\delta_0$ and cylinder density $\rho_\textrm{c}=10\rho$. The particle experiences a small rotation that is not visually observable in (c) and (d).} (e) Normalized gap thickness versus the normalized transverse displacement of the cylinder. 
Inset shows the trajectory of the Janus cylinder over a larger spatial extent.
 \label{fig:obs}
\end{figure}

In this Letter, we study lubrication forces when slippage \cite{lauga_noslip2007} properties change along the contacting surfaces. We consider a model of spatially varying slip length $\ell$ 
at either the surface of a flat wall or the surface of a cylinder (Fig.~\ref{fig:obs}a,b). Here, $\ell$ is defined as an effective property of the interface at some coarse-grained level. The slip length can be considered as a mesoscopic model emerging from small-scale features such as surface charges \cite{bocquet2007}, 
wall roughness \cite{ugis_2017}, superhydrophobicity \cite{Rothstein_ARFM}, liquid infusion \cite{wong_slips}, and temperature or solute concentration gradients \cite{Anderson1989, Ajdari_Bocquet_PRL_2006}.  

Using analytical and numerical treatments, we demonstrate that surface inhomogeneities give rise to particle trajectories such as oscillations, migration and propulsion. 
Underlying these phenomena is a normal lift force that arises from spatial variations in surface slippage. To illustrate this, consider the two configurations in Fig.~\ref{fig:obs}(c,d).
Both cases involve a cylinder of radius $r$ located a distance $\delta_0$ from a flat wall and immersed in a  fluid with viscosity $\eta$ and density $\rho$.  We assume small Reynolds number, $Re=\rho V r/\eta\ll 1$, where $V$ is the characteristic velocity of the cylinder. 

Figure \ref{fig:obs}(e) (blue) shows the trajectory of the cylinder falling freely under gravity next to a wall that has a single slip transition.  The trajectory is obtained from numerical simulations of Stokes equations 
coupled to Newton's equation of motion for the cylinder (see SI). As the cylinder passes the transition line, it migrates away from the wall a distance $\Delta$, that is comparable to $\delta_0$. In contrast, a wall with homogeneous slippage produces zero lift force (and consequently no wall-normal motion) on a cylinder \cite{kaynan}. Therefore, the lift arises here from the sudden change in slip length at the wall. By carrying out an expansion in the dimensionless slip length $L=\ell/\delta_0$, we will show that the lift force per unit length of the cylinder scales as $F_z \sim  \eta V^{\parallel} \varepsilon^{-1}L $ at the transition line, and that $\Delta \sim \ell$, where $\varepsilon=\delta_0/r\ll 1$. Importantly, 
this lift force can be comparable in magnitude to other lubrication forces. As an example, for a red blood cell traveling near glycocalyx 
\footnote{Typical parameters are: red blood cell  radius $r = \SI{3}{\micro\meter}$, cell-wall gap $\delta = \SI{0.5}{\micro\meter}$, fluid viscosity $\eta = \SI{1.5}{\milli\pascal\second}$, and speed $V^{\parallel} = \SI{0.286}{\milli\meter\per\second}$ estimated by multiplying the $\SI{100}{\per\second}$ shear rate and the cell radius \cite{davies2018elastohydrodynamic}},
variations of the slip length as small as a few nanometers  induce a lift force comparable to  the elastohydrodynamic one ($\sim 0.1$ pN) caused by glycocalyx deformation \cite{davies2018elastohydrodynamic}.

Figure~\ref{fig:obs}(e) (red) shows the trajectory (obtained from numerical simulations, see SI) of a Janus cylinder falling near a flat wall (Fig.~\ref{fig:obs}d). 
We observe a persistent normal drift along the trajectory, since the transition from slip to no-slip is now located on the traveling cylinder itself thus constantly inducing a wall-normal force. Scaling estimates, that will be obtained below, indicate that $\Delta \sim x_c\ell/l_\textrm{c}$,  where $l_\textrm{c} = \sqrt{2\delta_0 r}$ is the lubrication contact length and $x_c$ is the transverse displacement. As $\Delta$ becomes larger than $r$, one expects a saturation of normal migration, since lubrication forces become negligible in the bulk. Nevertheless, as shown in the inset of Fig.~\ref{fig:obs}(e), the effect holds for $\delta$ substantially larger than $r$.

\begin{figure}
 \centering
  \includegraphics[width=0.97\columnwidth]{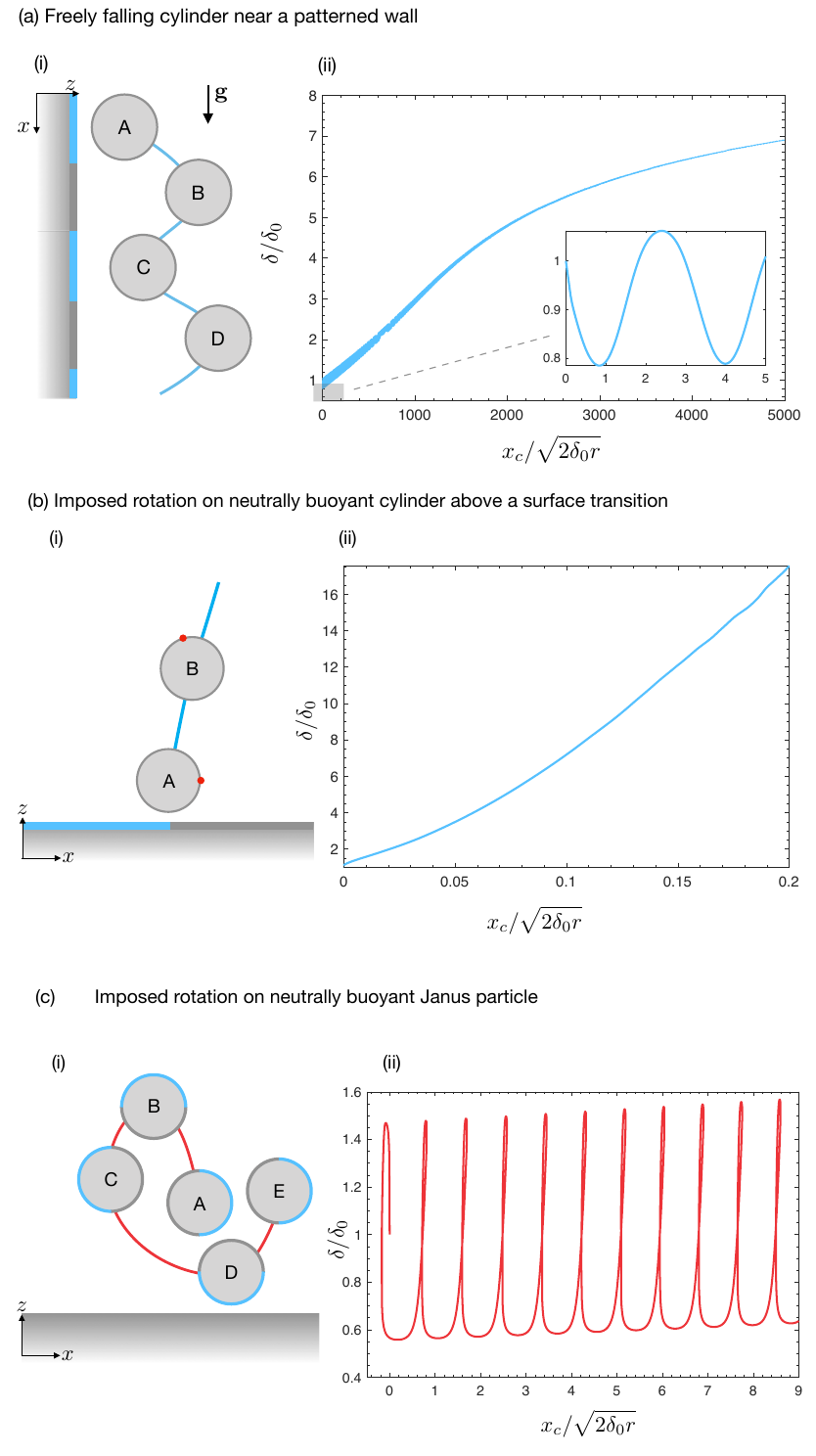}
 \caption[]{Cylinder trajectories induced by variation of slippage. (a) Cylinder falling near a wall that alternates between slip (blue) and no slip (grey). (b) Neutrally buoyant cylinder rotating above a slip-to-no-slip transition. (c) Neutrally buoyant Janus cylinder rotating next to a wall. Right column (ii) shows the normalized gap thickness as a function of the normalized transverse displacement.  For all configurations, $\delta_0/r = 0.05$ and $L = 1.79$. }
 \label{fig:trajecotry2}
\end{figure}

The coupling of rigid-body motion to slippage inhomogeneities can result in unexpected particle dynamics (Fig.~\ref{fig:trajecotry2}). 
In the following, we will study in detail these motions using lubrication theory and scaling laws.
At low Reynolds numbers, the force per unit length, $\textbf{F}=(F_x,F_z)$, and torque per unit length, $T$, on the cylinder are linearly related to the velocity $\textbf{V}=(V^{\parallel}, V^{\bot})$ and the angular speed $\Omega$ of the cylinder. This is expressed by the symmetric resistance matrix \cite{leal2007advanced}, 
\begin{align}
\begin{bmatrix}
F_x \\
F_z \\
T
\end{bmatrix}
=
-\eta
\begin{bmatrix}
    f_x^{\parallel} & -f_x^{\bot} &  r f_x^{\omega}\\
   -f_z^{\parallel} &  f_z^{\bot} & -r f_z^{\omega}\\
   r t^{\parallel} & -r t^{\bot} &  r^2 t^{\omega}
\end{bmatrix}
\begin{bmatrix}
V^{\parallel}\\
V^{\bot}\\
\Omega
\end{bmatrix}, \label{eq:resist-mat}
\end{align}
where $f_x^\bot=f_z^\parallel$, $t^\parallel=f_x^\omega$ and $t^{\bot}=f_z^\omega$.
\begin{figure}
 \centering
 \includegraphics[width=0.95\columnwidth]{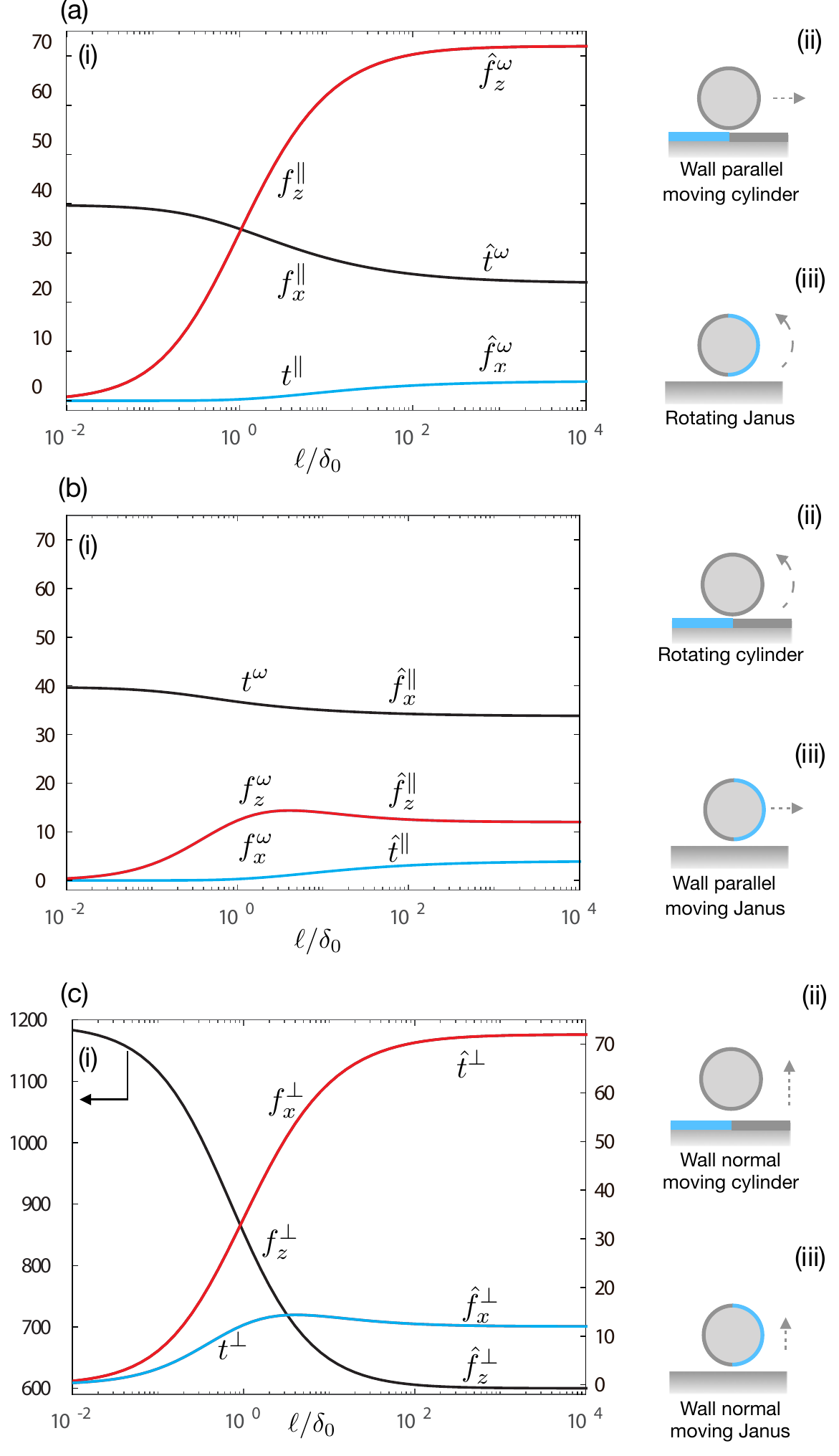}
\caption{Analytical resistance coefficients for both model systems ($\varepsilon = 0.05$). In the left column (i), the coefficients with no $ \hat{} $ refer to
configuration (ii) and the ones with $ \hat{} $ refer to 
(iii). 
}
 \label{fig:ResCoef}
\end{figure}
Assuming a small gap, $\varepsilon=\delta_0/r\ll 1$, we  explain below the procedure to determine the elements of the resistance matrix for the cylinder translating parallel to the wall with a slip-to-no-slip transition (Fig.~\ref{fig:ResCoef}a,ii). 

We non-dimensionalize the variables as,
\begin{align*}
     x &= l_\textrm{c}\,X, & z &= \delta_0\, Z, & h &= \delta_0\, H, \\
     u &= V^{\parallel}\, U, & w &= \frac{\varepsilon^{1/2} V^{\parallel}}{\sqrt{2}}\, W, & p &= \frac{l_\textrm{c}\eta V^{\parallel}}{\delta_0^{2}}\,P,
\end{align*}
where, $u,w$ are, respectively, the transverse and normal components of the fluid velocity, $p$ is the fluid excess pressure with respect to the atmospheric one. The cylinder surface is approximated as $h(x) = \delta_0 + x^2/(2r)$. Inserting the dimensionless variables into the continuity and Stokes equations and neglecting $\mathcal{O}(\varepsilon)$ terms, we obtain 
\begin{align}
\partial_X P =\partial_{ZZ} U, \ \ \ \ \partial_Z P = 0, \ \ \ \ 
\partial_X U + \partial_Z W = 0. \label{eq:reynolds}
\end{align}
In the laboratory frame of reference, the boundary conditions are,
\begin{align}
 &\left. W \right |_{Z=0,H} = 0, \qquad  \left.  U \right |_{Z=H} = 1, \\
&  \left. U \right |_{Z=0}  =  \left.  L \partial_Z U\right |_{Z=0} S.
  \label{eq:boundary cond}
 \end{align}
Equation \eqref{eq:boundary cond} accounts for slippage and is modelled through a Navier boundary condition. 
Moreover, $S$ equals one for $X<0$ and zero for $X>0$, where $X=0$ is the location of the transition from slip to no slip. The solution of Eqs.~\eqref{eq:reynolds} and \eqref{eq:boundary cond} is a combination of Couette and Poiseuille flows (see SI). From that solution, the fluid stress is projected onto the cylinder surface to obtain
\begin{align}
  f_x^{\parallel} &=  \sqrt{2} \varepsilon^{-1/2}\int_{-\infty}^{\infty}{\left(2XP + \partial_Z U \right)\textrm{d}X}, \label{eq:fxp}\\
  f_z^{\parallel} &= 2 \varepsilon^{-1} \int_{-\infty}^{\infty}{P \textrm{d}X},\label{eq:fz0} \\
  t^{\parallel} &=  \sqrt{2} \varepsilon^{-1/2}\int_{-\infty}^{\infty}{\partial_Z U \textrm{d}X}.  \label{eq:fz}
\end{align}
Figure \ref{fig:ResCoef}(a,i) shows the elements (Eqs.~\ref{eq:fxp}-\ref{eq:fz}) of the resistance matrix as a function of dimensionless slip length. When $\ell/\delta_0\rightarrow 0$, only the drag coefficient ($f_x^\parallel$) is non-zero, in agreement with the results for no slip surfaces \cite{jeffery}. 
For larger $\ell/\delta_0$, we note the emergence of non-zero elements related to lift force ($f_z^\parallel$)  (as reported numerically in Fig.~\ref{fig:obs}e) and torque  $ t^{\parallel}$. 

Figure \ref{fig:ResCoef}(b,i) shows the elements  (denoted by the $\hat{}$ symbol) of a Janus cylinder that translates parallel to a wall. We again observe the emergence of off-diagonal terms of the resistance matrix for $\ell\approx \delta_0$; in particular the lift force ($\hat f_z^\parallel$) causing the constant normal migration shown in Fig.~\ref{fig:obs}(e). The complete set of elements for both model systems is reported in Fig.~\ref{fig:ResCoef}. Note that due to the Lorentz reciprocal theorem \cite{leal2007advanced} there is a symmetry between the two model systems.

\begin{figure}
 \centering
 \includegraphics[width=0.95\columnwidth]{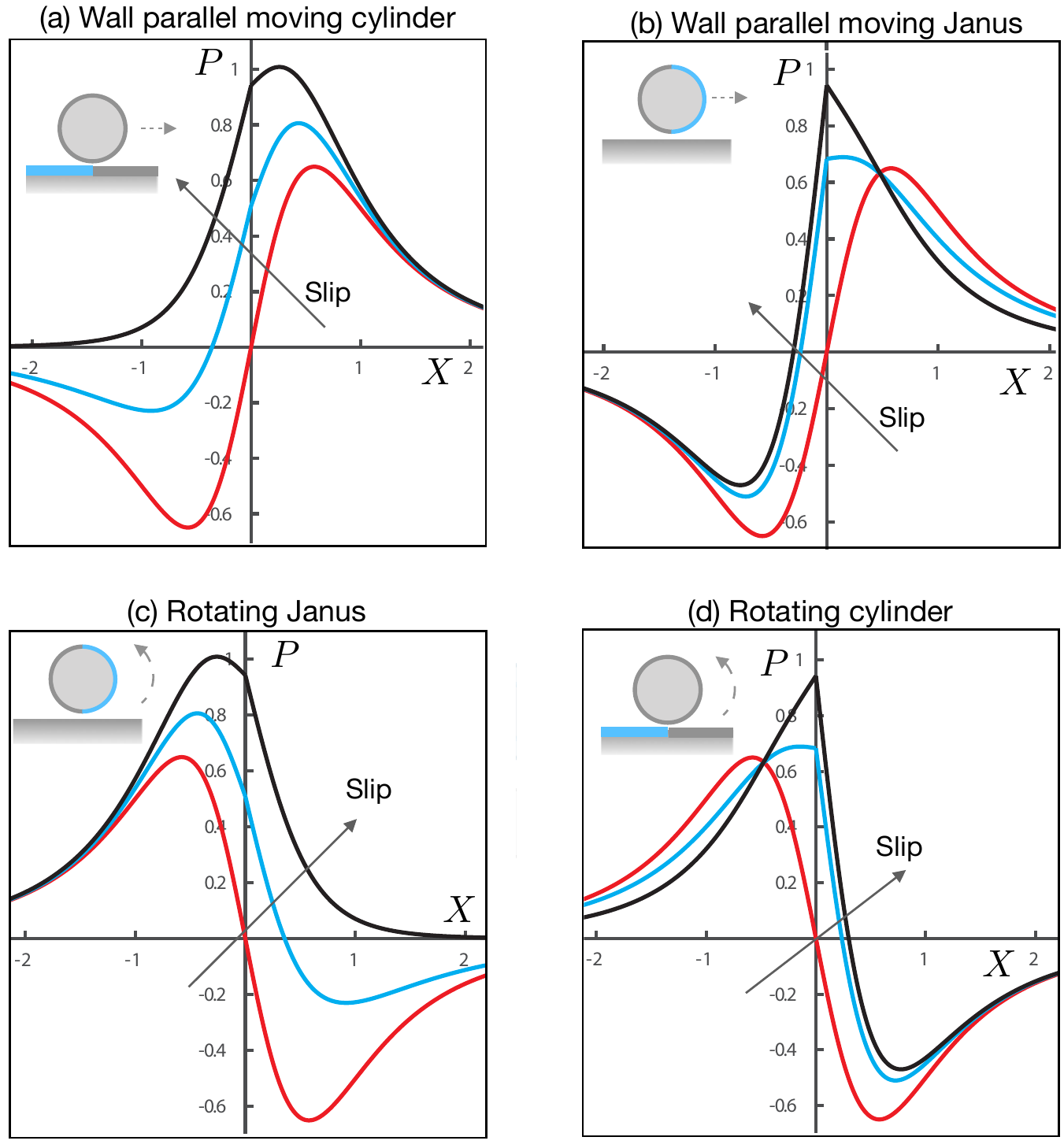}
\caption{Lubrication pressure distributions in four configurations and for three slip lengths,  $L = 0$ (red), $L = 1$ (blue) and $L = \infty$ (black).}
 \label{fig:Pressure}
\end{figure}

To understand in  detail the lift-induced mechanism of slip-to-no-slip transitions, we study the gap pressure distributions for three different slip lengths. Figure \ref{fig:Pressure}(a) shows the pressure distribution for a moving cylinder over an inhomogeneous wall (trajectory in Fig.~\ref{fig:obs}c). The no-slip solution maintains an anti-symmetric pressure distribution (red). The introduction of a finite slip length breaks this symmetry (blue and black), as the gap pressure necessary to accelerate the flow through the gap (with varying thickness) is increased over the slippery section ($X<0$).  Figure \ref{fig:Pressure}(b) shows the gap pressure of the wall-parallel moving Janus particle (trajectory in Fig.~\ref{fig:obs}d). Here, the induced lubrication pressure needs to accommodate both for varying gap thickness and for varying fluid shear in the gap, which results in a pressure peak over the slippery section of the Janus cylinder. Note that due to the symmetry between the two configurations the pressure distributions in (a) and (b) are the mirror of distributions (c) and (d), respectively. 

We now explain the trajectories in Fig.~\ref{fig:trajecotry2} using the  components of the resistance matrix shown in Fig.~\ref{fig:ResCoef}. 
 In Fig.~\ref{fig:trajecotry2}(a), a slip-to-no-slip transition pushes the cylinder away from the wall (Fig.~\ref{fig:obs}c), whereas a no-slip-to-slip transition produces a negative $F_z$, pulling the cylinder towards the wall. The  long-term net migration away from the wall (Fig.~\ref{fig:trajecotry2}a,ii) is partially  due to the fact that the push force is slightly larger than the pull force for each period due to different gap thicknesses during the pull and push events (see SI).

The second example involves a rotating neutrally buoyant cylinder (Fig.~\ref{fig:trajecotry2}b). 
When the cylinder is released above a slippage transition on the wall, we observe a migration in both $x-$ and $z-$directions  (see Fig.~\ref{fig:trajecotry2}b,ii).  The resistance coefficients in Fig.~\ref{fig:ResCoef}(b) and (c) explain this behavior. The imposed rotation produces a lift force ($F_z\sim\eta r\Omega f_z^\omega> 0$) and a negative transverse thrust ($F_x\sim-\eta r\Omega f_x^\omega<0$). However, as the cylinder migrates away from the wall, we have $V^{\bot}>0$, which leads to a positive transverse thrust ($F_x\sim\eta V^{\bot} f_x^\bot>0$). The $V^{\bot}$-generated thrust dominates over the $\Omega$-generated thrust, such that the cylinder moves in the positive $x$ direction.

\begin{table}
  \caption{Scaling laws for lift force ($F_z$) and wall-normal displacement ($\Delta$) for ST Wall or Janus particle. The imposed motion is either wall-parallel velocity or rotation. For the latter, the displacement shown is for one revolution.}
  \begin{tabular}{l c cc}
    \hline \hline
     & Motion &\hspace{0.5cm} Lift force\hspace{0.5cm} & Displacement \\
    \hline
    ST Wall & $V^{\parallel}$ &  $F_z \sim \eta V^{\parallel} \varepsilon^{-1} L$ &  $\Delta \sim \ell$  \\
    Janus & $V^{\parallel}$ &  $F_z \sim \eta V^{\parallel} \varepsilon^{-1} L$ &  $\Delta \sim \ell x_{\textrm{c}}/l_{\textrm{c}}$ \\
    ST Wall & $\Omega$ &  $F_z \sim \eta r \Omega \varepsilon^{-1} L$  &  $\Delta \sim   \sqrt{r/\delta} \ell $  \\
      Janus & $\Omega$ &  $F_z \sim \eta r \Omega \varepsilon^{-1} L$  &  $\Delta \sim \ell \sqrt{r/\delta}-\frac{\ell \sqrt{r/\delta}}{\sqrt{1+\ell r^{1/2} \delta^{-3/2}}}$
\end{tabular}
  \label{table:scaling}
\end{table}

The final example involves a rotating neutrally buoyant Janus cylinder (Fig.~\ref{fig:trajecotry2}c). The cylinder undergoes a spiralling motion that results in positive transverse propulsion and positive wall-normal migration. We  explain the  motion in  stages A-E depicted in Fig.~\ref{fig:trajecotry2}(c).
In stage $A$, the cylinder migrates upward since $\hat f_z^\omega>0$. It simultaneously migrates to the left since, due to the short exposure, $V^{\bot}$ is relatively small and we have $\lvert V^{\bot}\hat{f}_x^{\bot} \rvert < \lvert r\Omega \hat{f}_x^{\omega} \rvert$.
In  stage $B$, the Janus cylinder has rotated  such that no slippage remains in the gap.  Therefore, a lift force is no longer generated and the wall-normal drag force $\hat f_z^\bot$ hinders further upward migration. 
Stages $C$ and $D$ are the mirror of $A$ and $B$, respectively. Consequently, we observe a migration in the negative $z$ and positive $x$ directions.
As the cylinder reaches stage $E$, it has experienced a net translation in the positive $x$ direction and a small net migration away from the wall from its initial position $A$. This is due to the difference in the magnitude of fluid stresses in the gap when slip ($B$) and no slip ($D$) surfaces face the wall.

Finally, we turn to scaling analysis to estimate the normal displacement induced by the lift force. We focus on the instant where a cylinder is located above the transition from slip to no slip on the wall (Fig.~\ref{fig:obs}a). We assume no rotation, constant transverse and wall-normal velocities $V^\parallel$ and $V^\perp$, and negligible inertial effects (see SI). A wall-normal force balance yields $ \eta f_z^\parallel V^\parallel = \eta f_z^\bot V^\bot$, corresponding to the magnitude of the lift and drag components, respectively. By carrying out an expansion in $L\ll 1$ of $f_z^\parallel$ and $f_z^\bot$, we can approximate the two coefficients using
leading order terms as
\begin{equation}
  f_z^{\parallel} \simeq 4 \ell r/\delta_0^2 \quad \mbox{and} \quad f_z^{\bot} \simeq -3\sqrt{2}\pi \left(\delta_0/r\right)^{-3/2}.
  \label{eq:fz-force}
\end{equation}
By assuming that the lift force approximately acts over $l_\textrm{c}$, we have
\begin{equation}
V^{\parallel} = l_\textrm{c} / \tau \qquad \mbox{and} \qquad V^{\bot} = \Delta / \tau,
\label{eq:sc-est-vel}
\end{equation}
where $\tau$ is the time it takes to traverse $l_\textrm{c}$, and $\Delta$ is the displacement from the wall. Inserting these estimates in the wall-normal force balance yields  $\Delta \sim \ell$. This is in agreement with numerical results shown Fig.~\ref{fig:obs}(c), where we observe $\Delta \approx \delta_0$ for $\ell \approx \delta_0$. 
For a Janus cylinder (Fig.~\ref{fig:obs}b), there is no fixed time over which the displacement occurs, because the transition point travels along with the cylinder. Therefore, we expect a displacement $\sim \ell$ for each $l_\textrm{c}$ traversed. This yields, 
\begin{equation}
\Delta \sim \ell x_{\textrm{c}}/l_\textrm{c} ,
\end{equation}
where $x_{\textrm{c}}$ is the distance traversed along the wall.
Table~\ref{table:scaling}, summarizes the scaling estimates of the lift force and normal displacement, induced by transverse motion or rotation, for both systems in Fig.~\ref{fig:obs}(a,b) (see SI).

To conclude, we have used a combination of numerical simulations, analytical results obtained from lubrication theory, and scaling arguments, to describe how fluid-immersed objects approaching contact with spatially inhomogeneous slippage properties may encounter non-trivial emergent forces and torques. In the presence of small slippage inhomogeneities, our scaling estimate suggests that the lift force, $F_{z,\textrm{slip}}$, has to be considered alongside  other lubrication forces.
For instance, using atomic-force microscopy, the elastohydrodynamic force was reported~\cite{2020_PRL_Zhang_Salez_AFM} to be $F_{z,\textrm{EHD}}\sim$\SI{}{\nano \newton}. Considering the same experimental setup, but with a rigid substrate exhibiting a $\sim 1$ \SI{}{\nano \meter} slip inhomogeneity, we obtain $F_{z,\textrm{slip}}\sim$\SI{}{\nano \newton}, 
 i.e.~the same order of magnitude as for the soft substrate
 \footnote{Experimental parameters from \cite{2020_PRL_Zhang_Salez_AFM}: radius $r = \SI{60}{\micro\meter}$, probe-wall gap $\delta = \SI{50}{\nano\meter}$, fluid viscosity $\eta = \SI{96}{\milli\pascal\second}$, and speed $V^{\parallel} = \SI{0.57}{\milli\meter\per\second}$. Inserting these values in $F_z$ in Tab.~\ref{table:scaling} (1st row) gives $F_{z,\textrm{slip}}=F_z l_c\sim 1$ \SI{}{\nano \newton}. Here, $l_c=\sqrt{2R\delta}$ is the relevant length scale for  the lift force.}. Such  slippage inhomogeneities may arise from small variations in surface preparation, localized defects, or cleaning processes. For example, in the experiments of a traveling Janus particle near a wall \cite{Ketzetzi_PRL_2020}, variations of contact angle up to 10$^\circ$ on the same substrate were reported.
Finally, our numerical results show how non-trivial particle trajectories can spontaneously emerge from slippage inhomogeneities. This opens up new interesting opportunities for the design of interfaces to control and influence nearby particle motion, as well as to reduce friction and wear. It also provides a foundation to explore more realistic situations found in nature and especially biology.

\begin{acknowledgements}
A.R., S.B. and U.L. were funded through the Knut and Alice Wallenberg Foundation
(KAW 2016.0255), the Swedish Foundation for Strategic Research (SSF, FFL15:0001) and Swedish Research Council (VR-2014-5680). We also thank Abdelhamid Maali and Juho Lintuvuori for valuable comments on the manuscript.
\end{acknowledgements}
\bibliography{CylinderMotionNearWall}
\end{document}


\title{Supplemental Material to ``Lift induced by slip inhomogeneities in lubricated
contacts''}%

\author{Aidan Rinehart$^1$}%
\author{U\v{g}is L\={a}cis$^1$}
\author{Thomas Salez$^{2,3}$}
\author{Shervin Bagheri$^1$}

\email[corresponding author:]{ shervin@mech.kth.se}
\affiliation{$^1$Department of Mechanics, KTH Royal Institute of Technology, Stockholm 11428, SE}
\affiliation{$^2$ Univ. Bordeaux, CNRS, LOMA, UMR 5798, F-33405, Talence, France.} 
\affiliation{$^3$ Global Station for Soft Matter, Global Institution for Collaborative Research and Education, Hokkaido University, Sapporo, Hokkaido 060-0808, Japan.}

\date{Nobember 2019}%

\maketitle

In this material we provide more detailed
analytical derivations of lubrication forces in sections \ref{sec:derive1} and \ref{sec:derive2}. In section~\ref{sec:scale} we provide Taylor expansions of force coefficients needed for displacement scaling estimates and the details of scaling law derivations.
Then, in section~\ref{sec:scale-correct} we discuss a correction of the
single transition scaling law that allows to capture the net migration away from the
patterned wall.
Finally, in section~\ref{sec:num-meth} we provide details of the numerical method
used for producing the particle trajectories.

\section{Analytical derivations of forces and resistance coefficients} \label{sec:derive1}
In the following we provide some detailed derivations for the analytical expressions and scaling arguments provided in the main text.  We begin with the forces and torques generated by the fluid resisting the motion of the cylinder,  
\begin{align}
  & \bm{F} = \int_S{\bm{\sigma}\cdot\bm{n}}\textrm{d}S, & T = \int_S{\bm{r}\times\left(\bm{\sigma}\cdot\bm{n}\right)}\textrm{d}S.
\end{align}
Here, $S$ is the cylinder surface, $\bm{\sigma}$ is the fluid stress tensor, $\bm{n}$ the surface normal vector, and $\bm{r}$ the cylinder radius. The forces and torques expressed in terms of the flow velocities, $u$, $w$ and pressure $p$ are as follows, 
\begin{align}
   F_x = \int_S{\left( -p + 2\eta \frac{\partial u}{\partial x}\right)n_x + \eta\left(\frac{\partial u}{\partial z} + \frac{\partial w}{\partial x} \right)n_z}\textrm{d}S \label{eq:Fx}, \ \ \ \ F_z =  \int_S{\left( -p + 2\eta \frac{\partial w}{\partial z}\right)n_z + \eta\left(\frac{\partial u}{\partial z} + \frac{\partial w}{\partial x} \right)n_x}\textrm{d}S, \\
   T = r\int_S{2\eta\left(\frac{\partial w}{\partial z}-\frac{\partial u}{\partial x}\right)n_x n_z + \eta\left(\frac{\partial u}{\partial z} + \frac{\partial w}{\partial x} \right)\left(n_x^2 - n_z^2\right)}\textrm{d}S. \label{eq:T}
\end{align}
We proceed with the exact derivations of the force and torques for the three modes of motion; wall parallel, wall normal and rotation, additionally showing that the resistance matrix obtains the expected symmetric form. 
\subsection{Wall Parallel and Rotation Motion}    
The wall parallel and rotation motions share the non-dimensional relationships obtained from the standard lubrication scaling given in the main text. The following derivation also requires the non-dimensional unit normal vector $\bm{N} = [\sqrt{2}\varepsilon^{1/2}X\hat{e}_x, -\hat{e}_z]$. Introduction of these non-dimensional variables into the Eqs. \ref{eq:Fx}-\ref{eq:T} provide force and torque expressions with the scale separation parameter, $\varepsilon = \delta_0/r$. Application of the lubrication scaling to Eq. \ref{eq:Fx} gives,  
\begin{align}
  &F_x =  \int_{-\infty}^{\infty}{\left( -\sqrt{2}\eta V\varepsilon^{-1/2}XP + 2\sqrt{2}\eta V\varepsilon^{1/2}X\frac{\partial U}{\partial X}\right) - \eta\left(\sqrt{2}\varepsilon^{-1/2}V\frac{\partial U}{\partial Z} + \varepsilon^{1/2}\frac{\partial W}{\partial X} \right)}\textrm{d}X.
  \end{align}
The leading order terms of the non-dimensional integral are grouped into the leading order expression,
  \begin{align}
  &F_x = -\sqrt{2}\eta V \varepsilon^{-1/2} \int_{-\infty}^{\infty}{2XP + \frac{\partial U}{\partial Z}}\textrm{d}X. \label{eq:fx}
  \end{align}
This is the form we report in the main text Eq. 4, with viscosity and velocity scaled out consistent with resistance matrix notation. Moving to the force along the z-axis one obtains,
  \begin{align}
  &F_z =  \int_{-\infty}^{\infty}{\left( 2\eta V \varepsilon^{-1}P - 2\sqrt{2}\frac{\partial W}{\partial Z}\right) + \eta\left(2V\frac{\partial U}{\partial Z} + \sqrt{2}V\frac{\partial W}{\partial X} \right)X}\textrm{d}X.
  \end{align}
Considering only the leading order terms of $\varepsilon$ one obtains,
  \begin{align}
  &F_z = 2\eta V \varepsilon^{-1}\int_{-\infty}^{\infty}{P}\textrm{d}X. \label{eq:fz}
 \end{align}
This is the wall normal force form presented in the main text Eq. 5, with viscosity and velocity factored out. Finally torque for wall parallel motion and rotation takes the form,
 \begin{align}
    &T = r\eta V\int_{-\infty}^{\infty}{-2\sqrt{2} \varepsilon^{1/2}X\left(\frac{\partial U}{\partial X}+\frac{\partial W}{\partial Z} \right) - \sqrt{2} \left(\varepsilon^{-1/2}\frac{\partial U}{\partial Z} + \frac{\varepsilon^{1/2}}{2}\frac{\partial W}{\partial X} \right)}
       - 2 X^2\left(\sqrt{2}\varepsilon^{1/2}\frac{\partial U}{\partial Z} + \varepsilon^{3/2} \frac{\partial W}{\partial X}\right)\textrm{d}X.
\end{align}
Lastly we obtain Eq. \ref{eq:ty} the leading order equation for torque presented in the main text as Eq. 5, 
\begin{align}
    &T = -\sqrt{2} \eta V\varepsilon^{-1/2}r \int_{-\infty}^{\infty}{\frac{\partial U}{\partial Z}}\textrm{d}X. \label{eq:ty}
\end{align}
One can note the different powers of $\varepsilon^{\alpha}$ where $F_x$ and $T$ have $\alpha = -1/2$ and $F_z$ has $\alpha = -1$. Consequently the magnitude of these forces and torques will grow at different rates as the cylinder-wall gap varies.

\subsection{Wall Normal Motion}
Wall normal motion takes a different non-dimensional lubrication scaling relationships due to the reference velocity taken in the wall normal direction. The velocity and pressure terms take the following form in the wall normal direction,   
\begin{align*}
    u &= \sqrt{2}\varepsilon^{-1/2}VU, & w &= VW, & p &= \frac{2\eta V}{r\varepsilon^{2}}P.
  \end{align*}
Introducing the above non-dimensional terms into Eq. \ref{eq:Fx} yields the following equation, 
\begin{align}    
    &F_x = \int_{-\infty}^{\infty}{\left( -4\eta V\varepsilon^{-1}XP + 4\eta\varepsilon^{-1/2}VX \frac{\partial U}{\partial X}\right) - \eta\left(2\varepsilon^{-1}V\frac{\partial U}{\partial Z} +  V\frac{\partial W}{\partial X} \right)}\textrm{d}X.
\end{align}
Gathering the leading order terms of $\varepsilon$ gives the force along the x-axis,  
\begin{align}
    &F_x = -2\eta V \varepsilon^{-1}\int_{-\infty}^{\infty}{2XP + \frac{\partial U}{\partial Z}}\textrm{d}X. \label{eq:fx2}
\end{align}
The only difference from the wall parallel and rotation x-axis forcing, Eq. \ref{eq:fx}, is a higher power of $\varepsilon$ and a factor of $\sqrt{2}$. The force in the z direction obtains the following form, 
\begin{align}
    &F_z =  \int_{-\infty}^{\infty}{2\sqrt{2}\eta V \varepsilon^{-3/2}P - 2\sqrt{2}\eta V \varepsilon^{-1/2}\frac{\partial W}{\partial Z} + \eta V\left(2\sqrt{2}\varepsilon^{-1/2}\frac{\partial U}{\partial Z} + \sqrt{2}\varepsilon^{1/2}\frac{\partial W}{\partial X}\right)}\textrm{d}X.
\end{align}
The leading order equation in $\varepsilon$ is,    
\begin{align}
    &F_z =  2\sqrt{2}\eta V \varepsilon^{-3/2}\int_{-\infty}^{\infty}{P}\textrm{d}X. \label{eq:fz2}
\end{align}
The force in the z direction for the wall normal motion has the largest power of $\varepsilon$ with $\alpha = -3/2$. Following the same procedure the wall normal motion generates a torque,  
\begin{align}
    &T = r\int_{-\infty}^{\infty}{4\eta V\frac{\partial U}{\partial Z} - \eta \left( 2V \varepsilon^{-1}\frac{\partial U}{\partial Z} + V \frac{\partial W}{\partial X}\right) - 4\eta V\frac{\partial W}{\partial Z} + \eta VX^2 \left( 4\frac{\partial U}{\partial Z} + 2\varepsilon\frac{\partial W}{\partial X} \right)}\textrm{d}X.
\end{align}
Gathering the leading order terms of $\varepsilon$ produces the torque equation,    
\begin{align}
    &T = -2\eta r V \varepsilon^{-1}\int_{-\infty}^{\infty}{\frac{\partial U}{\partial Z}}dX.\label{eq:ty2}
\end{align}
The complete description of the forces and torques outlined thus far require solutions for the pressure and velocity fields provided in the following section.

\section{Solutions to Reynolds Equations} \label{sec:derive2}
The pressure and velocity solutions are obtained by solving the Reynolds equation introduced in the main text as Eq. 2. The Reynolds equation can be solved subject to three sets of boundary conditions corresponding to the three modes of motion. The boundary conditions for the inhomogeneous slip wall and motion combinations are summarized in Table \ref{table:reynolds bc}. 
\begin{table}
  \caption{Reynolds equation boundary conditions}
  \begin{tabular}{c | c c c c c c}
    \hline \hline
     & $U(X,H)$ & $U(X<0,0)$ & $U(X>0,0)$ & $W(X,H)$ & $W(X,0)$ & $P(\pm \infty)$  \\
    \hline
    $V^{\parallel}$ & 1 & $L\partial_Z U$ & 0 & 0  & 0 & 0  \\
    $V^{\bot}$      & 0 & $L\partial_Z U$ & 0 & 1  & 0 & 0  \\
    $\Omega$        & 1 & $L\partial_Z U$ & 0 & $2X$ & 0 & 0  \\
  \end{tabular}
  \label{table:reynolds bc}
  \end{table}
Reynolds equation solutions are obtained through first integrating the velocity, $U$. Then applying the boundary conditions gives relationships for $U$ as functions of geometry, $H$, and pressure, $P$. Next the continuity equation can be used to close the problem yielding an expression for pressure. As discussed in the main text the domain is split into two segments, $X<0$ and $X>0$, due to the variation in wall slip length. Therefore for every case there are two Reynolds equations that need to be solved. Lastly remaining unknown integration constants are obtained by matching the left and right domain pressure solution and mass flux at $X=0$. The solutions to the pressure and velocity fields at the the cylinder surface allow for the evaluation of the integrals in Eqs. \ref{eq:fx}, \ref{eq:fz}, \ref{eq:ty}, \ref{eq:fx2}, \ref{eq:fz2}, and \ref{eq:ty2}, producing the desired resistance coefficients. The resistance matrix of Eq. 1 is presented here in a slightly different form with $\varepsilon$ factored out so that the $R_{ii}$ represents the evaluation of the integral and constant, 
  \begin{align}
  \begin{bmatrix}
    F_x \\
    F_z \\
    T 
  \end{bmatrix}
  =
  -\eta
  \begin{bmatrix}
    \varepsilon^{-1/2}R_{11} & -\varepsilon^{-1}R_{12} & r\varepsilon^{-1/2}R_{13} \\
    -\varepsilon^{-1}R_{21} & \varepsilon^{-3/2}R_{22} & -r\varepsilon^{-1}R_{23} \\
    r\varepsilon^{-1/2}R_{31} & -r\varepsilon^{-1}R_{32} &  r^2\varepsilon^{-1/2}R_{33}  
  \end{bmatrix}
  \begin{bmatrix}
    V^{\parallel} \\
    V^{\bot} \\
    \Omega 
  \end{bmatrix}.
\end{align}
We now present the result of all nine resistance coefficient integrals. Starting with the wall parallel motion which produces the following resistance coefficients:
\begin{align}
&R_{11} = \sqrt{2}\int_{-\infty}^{\infty}{2XP + \frac{\partial U}{\partial Z}}dX = \frac{4 \sqrt{2} \left(-1+\sqrt{\frac{1}{1+4 L}}+L \left(-1+3 \sqrt{\frac{1}{1+4 L}}+L \left(3+5 \sqrt{\frac{1}{1+4 L}}\right)\right)\right) \pi }{-1+\sqrt{\frac{1}{1+4 L}}+2 L (1+5 L)}, \label{eq:R11}\\
  &R_{21} = 2\int_{-\infty}^{\infty}{P}dX = \frac{3 \left(4 L \left(1+3 L+\sqrt{\frac{1}{1+4 L}}\right)-(2+5 L) \ln(1+4 L)\right)}{-1+\sqrt{\frac{1}{1+4 L}}+2 L (1+5 L)},\\
  &R_{31} = \sqrt{2}\int_{-\infty}^{\infty}{\frac{\partial U}{\partial Z}}dX = \frac{2 \sqrt{2} \left(1-\sqrt{\frac{1}{1+4 L}}+L \left(3+L-5 \sqrt{\frac{1}{1+4 L}}-5 L \sqrt{\frac{1}{1+4 L}}\right)\right) \pi }{-1+\sqrt{\frac{1}{1+4 L}}+2 L (1+5 L)}.
  \end{align}
In the limit of $L \rightarrow 0$ one obtains the classic solution of Jeffrey \cite{jeffery}: $R_{11} = 2\sqrt{2}\pi$, $R_{21} = 0 $, and $R_{31} = 0$, and in the other limit as $L \rightarrow \infty$: $R_{11} = 6\sqrt{2}\pi/5$, $R_{21} = 18/5$, and $R_{31} = \sqrt{2}\pi/5$. For the wall normal motion the resistance coefficients are: 
\begin{align}
&R_{12} = -2\int_{-\infty}^{\infty}{2XP + \frac{\partial U}{\partial Z}}dX = R_{21},\\
\begin{split}
&R_{22}= -2\sqrt{2}\int_{-\infty}^{\infty}{P}dX =\\&\frac{\left( \pi^2\sqrt{\frac{1}{1+4 L}}( 18 +36L-132L^2-360L^3) +6 \left(-3+2 L^2 (8+5 L (4+5 L))\right)\pi ^2-9 (4 L (-1+2 L)+\ln(1+4 L))^2\right)}{\left(8 \sqrt{2} L^2 \left(-1+\sqrt{\frac{1}{1+4 L}}+2 L (1+5 L)\right) \pi \right)}, \\
  \end{split}\\
&R_{32} = -2\int_{-\infty}^{\infty}{\frac{\partial U}{\partial Z}}dX = \frac{3 \left(4 L \left(-3+L-\sqrt{\frac{1}{1+4 L}}\right)+(4+5 L) \ln(1+4 L)\right)}{2 \left(-1+\sqrt{\frac{1}{1+4 L}}+2 L (1+5 L)\right)}.
\end{align}
In the limit of $L \rightarrow 0$ one obtains the classic solution of Jeffrey: $R_{12} = 0$, $R_{22} = \frac{3\left(-48+25\pi^2\right)}{20\sqrt{2}\pi}$, and $R_{32} = 0$, in the other limit $L \rightarrow \infty$: $R_{12} = 18/5$, $R_{22} = 3\sqrt{2}\pi$, and $R_{32} = 3/5$. Finally for the rotation motion we obtain the following set of resistance coefficients:
\begin{align}
&R_{13} = \sqrt{2}\int_{-\infty}^{\infty}{2XP + \frac{\partial U}{\partial Z}}dX = R_{31},\\
&R_{23} = 2\int_{-\infty}^{\infty}{P}dX = R_{32},\\
&R_{33} = \sqrt{2}\int_{-\infty}^{\infty}{\frac{\partial U}{\partial Z}}dX = \frac{\sqrt{2} \left(L+7 L \sqrt{\frac{1}{1+4 L}}+4 \left(-1+\sqrt{\frac{1}{1+4 L}}\right)+L^2 \left(17+5 \sqrt{\frac{1}{1+4 L}}\right)\right) \pi }{-1+\sqrt{\frac{1}{1+4 L}}+2 L (1+5 L)}.\label{eq:R33}
\end{align}
In the limit of $L \rightarrow 0$ one obtains the classic solution of Jeffrey: $R_{13} = 0$, $R_{23} = 0$, and $R_{33} = 2\sqrt{2}\pi$, and in the other limit as $L \rightarrow \infty$: $R_{13} = \sqrt{2} \pi /5$, $R_{23} = 3/5 $, and $R_{33} = 17\sqrt{2}\pi/10$. Equations \ref{eq:R11}-\ref{eq:R33} verify the known symmetry of the resistance coefficient matrix, $R_{12}=R_{21}$, $R_{13} = R_{31}$, and $R_{23}=R_{32}$, reducing the overall resistance matrix to 6 unique terms.

The same process as detailed above can be repeated for the Janus cylinder with the only change being the boundary conditions. The Janus cylinder has slip occurring on the cylinder surface. These boundary conditions are summarized in Table \ref{table:reynolds bc janus}. 
\begin{table}
  \caption{Reynolds equation boundary conditions Janus cylinder}
  \begin{tabular}{c | c c c c c c}
    \hline \hline
     & $U(X<0,H)$ & $U(X>0,H)$ & $U(X,0)$ & $W(X,H)$ & $W(X,0)$ & $P(\pm \infty)$  \\
    \hline
    $V^{\parallel}$ & 1 & $L\partial_Z U$ & 0 & 0  & 0 & 0  \\
    $V^{\bot}$      & 0 & $L\partial_Z U$ & 0 & 1  & 0 & 0  \\
    $\Omega$        & 1 & $L\partial_Z U$ & 0 & $2X$ & 0 & 0  \\
  \end{tabular}
  \label{table:reynolds bc janus}
\end{table}
For brevity we summarize the results in the resistance matrix utilizing a transformation matrix relating the Janus cylinder terms to the inhomogeneous wall slip solutions, 
\begin{equation}
\begin{bmatrix}
\hat{f}_x^{\parallel} \\
\hat{f}_x^{\bot} \\
\hat{f}_x^{\omega} \\
\hat{f}_z^{\bot} \\
\hat{f}_z^{\omega} \\
\hat{t}^{\omega}
\end{bmatrix}
 = 
\begin{bmatrix}
0 & 0 & 0 & 0 & 0 & 1\\
0 & 0 & 0 & 0 & 1 & 0\\
0 & 0 & 1 & 0 & 0 & 0\\
0 & 0 & 0 & 1 & 0 & 0\\
0 & 1 & 0 & 0 & 0 & 0\\
1 & 0 & 0 & 0 & 0 & 0
\end{bmatrix}
\begin{bmatrix}
f_x^{\parallel} \\
f_x^{\bot} \\
f_x^{\omega} \\
f_z^{\bot} \\
f_z^{\omega} \\
t^{\omega}
\end{bmatrix}.
\label{eq:transform}
\end{equation}
We remind the reader that for the Janus cylinder the evaluation of Eqs. \ref{eq:R11}-\ref{eq:R33} will be consistent, but they will be associated with different integrals. For example the Janus cylinder has $\hat{R}_{11} = \sqrt{2}\int_{-\infty}^{\infty}{2XP+\partial_Z U \textrm{d}X}$ which is equal to inhomogeneous wall $R_{33} = \sqrt{2}\int_{-\infty}^{\infty}{\partial_Z U \textrm{d}X}$.

\section{Scaling Law Derivations} \label{sec:scale}
In the following we provide additional derivation details of the scaling laws presented in the main text. The Taylor expansion about $L =0$ for the wall normal lift and drag coefficients for the single transition wall respectively are as follows: 
\begin{align}
    f_z^{\parallel} = \varepsilon^{-1}\left(4 L-6 L^2+\frac{107 L^3}{10} + \mathcal{O}(L^4)\right), \label{eq:fzp}\\
     f_z^{\bot} = \varepsilon^{-3/2}\left(-3 \left(\sqrt{2} \pi \right)+\frac{9 \pi  L}{2\sqrt{2}}+\left(\frac{16 \sqrt{2}}{\pi }-\frac{45 \pi }{4 \sqrt{2}}\right)L^2 +\left(-\frac{76 \sqrt{2}}{\pi }+\frac{63 \pi }{2 \sqrt{2}}\right) L^3+ \mathcal{O}(L^4)\right). \label{eq:drag} 
\end{align}
The leading order coefficients for lift, $f_z^{\parallel} \approx 4 \varepsilon^{-1} L$, and drag, $f_z^{\bot} \approx -3\sqrt{2} \pi \varepsilon^{-3/2}$, are used to set wall normal forces equal and the wall normal displacement is obtained shown in the main text. For the Janus cylinder traveling parallel to the wall the Taylor expansion leading to Eq. 8 is as follows,
\begin{align}
    \hat{f}_z^{\parallel} = \varepsilon^{-1}\left(2 L-4 L^2+\frac{34 L^3}{5} + \mathcal{O}(L^4)\right), \label{eq:Tayl-exp-fzpar} \\ 
    \hat{f}_z^{\bot} = \varepsilon^{-3/2}\left(-3 \left(\sqrt{2} \pi \right)+\frac{9 \pi  L}{2 \sqrt{2}}+\left(\frac{16 \sqrt{2}}{\pi }-\frac{45 \pi }{4 \sqrt{2}}\right) L^2+\left(-\frac{76 \sqrt{2}}{\pi }+\frac{63 \pi }{2 \sqrt{2}}\right) L^3 + \mathcal{O}(L^4)\right). \label{eq:Jdrag}
\end{align}
Then we can approximate coefficients for lift as $\hat{f}_z^{\parallel} \approx 2 \varepsilon^{-1}L$, and drag as $\hat{f}_z^{\bot} \approx -3\sqrt{2}\pi \varepsilon^{-3/2}$. Balancing the approximate lift and drag forces provides the wall normal displacement estimate (see Table~1 in main paper), which is
obtained through neglecting inertial effects of the cylinder.

\begin{figure}
    \centering
    \includegraphics[width=0.5\columnwidth]{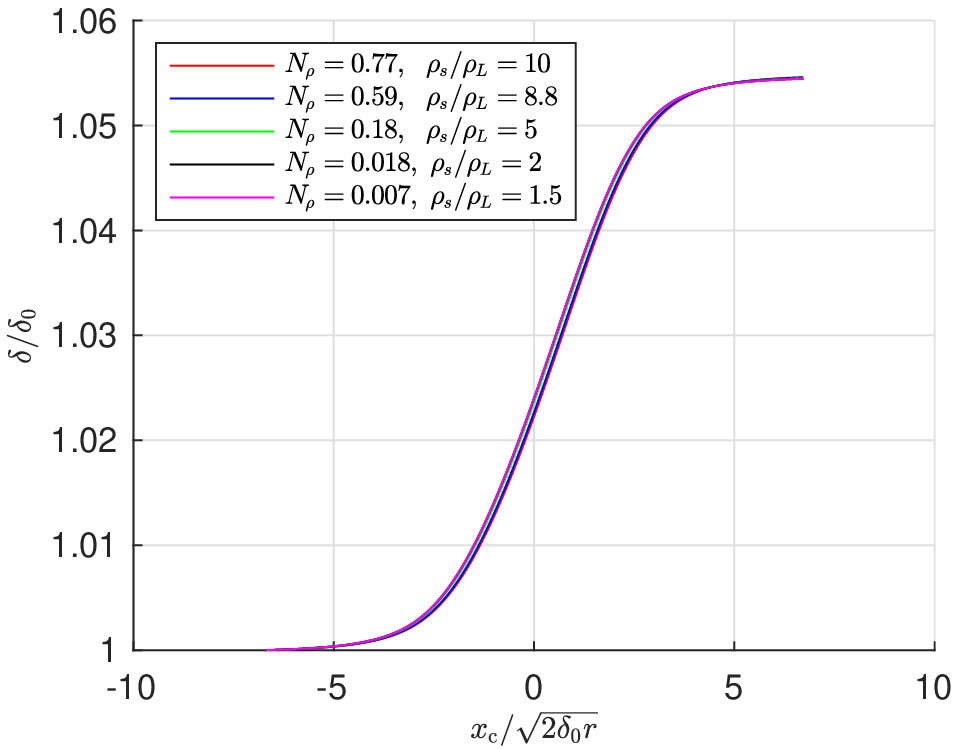}
    \caption{Wall normal displacements as a function of transverse position for several cylinder densities. We can observe no discernible difference in the trajectory and in the final displacement for the range of densities presented in agreement with the condition on $N_{\rho}$.}
    \label{fig:density}
\end{figure}

To elaborate on the importance of the cylinder inertia, we explain here the scaling estimate
derivations with cylinder inertia taken into account. The full Newton's equation of motion
for cylinder velocity away from the wall can be written as
\begin{equation}
m \frac{dV^\perp}{dt} = F_z^\parallel + F_z^\perp,
\end{equation}
where $F_z^\parallel$ is the slip-transition induced lift force per unit length, $F_z^\perp$ is the
standard drag force per unit length and $m$ is cylinder mass per length.
If either the cylinder mass or the change of wall-normal
velocity is sufficiently small -- the latter can occur if the inertial term acts only for a very
small interval in time, that is, cylinder accelerates to steady velocity quickly and
most of the motion happens in steady, non-inertial regime --,
the inertial term can be neglected and the overall wall-normal motion would be
governed by the balance of the lift and drag forces, as assumed in the main paper. To understand
if we fulfill one of these conditions, we can investigate the initial stage of the motion, where
inertia is always important (because there is acceleration) but wall-normal velocity
is very small (and thus
the drag force is very small). Assuming constant acceleration, we can balance the inertia
and lift force
\begin{equation}
m \frac{V^\perp}{\tau_a} \sim \eta f_z^\parallel V^\parallel, \label{eq:Newt-sec-law}
\end{equation}
where we have defined the acceleration time scale
$\tau_a$ over which the particle reaches the steady
wall-normal velocity $V^\perp$. Assuming that we reach the steady wall-normal velocity $V^\perp$
used in the inertia-less assumption $V^\perp = V^\parallel \Delta / l_c \sim V^\parallel l / l_c$
we get the following estimate of the acceleration time scale
\begin{equation}
\tau_a \sim \frac{m}{\eta} \epsilon^{3/2},
\end{equation}
where we have also used the Taylor expansion (\ref{eq:Tayl-exp-fzpar}) of coefficient $f_z^\parallel$.
Now, recall that the wall-normal motion away from the wall happens over transition time scale
$\tau = l_c / V^\parallel$, defined in the main paper.
In order to be able to neglect the inertial term, we require that the particle accelerates
to the steady wall-normal velocity in a time interval much smaller compared to the transition time,
i.e., $\tau_a \ll \tau$. This condition can be rewritten as
\begin{equation}
\frac{\tau_a}{\tau} \sim \frac{m\, \varepsilon\, V^{\parallel}}{\eta\, r} := N_\rho \ll 1.
\end{equation}
We have  validated
this condition in our numerical
experiments, by considering different density ratios. In Fig.~\ref{fig:density} we see that for time
scale ratios up to $N_\rho = 0.77$ and density ratios $\rho_c/\rho = 10$ the trajectories
of the cylinder are almost indistinguishable. This provides additional confirmation that the numerical
experiments provided in the main paper are relevant to density ratios found in biological applications.
For example, for red blood cell example we found $N_\rho = 10^{-8}$, which also shows that the
red blood cells are expected to follow the non-inertial scaling as derived in the main paper.

Next, following the negligible inertia assumption, we provide the derivation of the scaling laws for cylinder migration due to imposed rotation on neutrally buoyant cylinders. We begin with the inhomogeneous wall. We consider the net migration over one revolution. This provides us with cylinder velocities of $\Omega \sim \pi/\tau$ and $V^{\bot}\sim \Delta/\tau$, where $\tau$ is the time scale. The force coefficients are again approximated with a Taylor series expansion around $L = 0$ where the drag is still given by Eq. \ref{eq:drag} and the lift is given as,
\begin{align}
    f_z^{\omega} = \varepsilon^{-1}\left(2 L-4 L^2+\frac{34 L^3}{5} + \mathcal{O}(L^4)\right). 
\end{align}
This produces the leading order estimates as $f_z^{\omega} \approx 2 \varepsilon^{-1} L$ and $f_z^{\bot} \approx -3\sqrt{2}\pi\varepsilon^{-3/2}$. Then setting $f_z^{\omega} \eta r \Omega = f_z^{\bot}\eta V^{\bot}$ provides us with the scaling relationship for one revolution of a neutrally buoyant smooth cylinder above a inhomogeneous wall as,
\begin{equation}
    \Delta \sim \ell\sqrt{\frac{r}{\delta}}.
\end{equation}
To obtain a similar argument for how the Janus cylinder migrates away from the wall requires a slightly different approach as the force acting on the Janus cylinder will vary drastically as the cylinder rotates. We consider a simplification where migration is modeled in two steps; away from the wall for half a revolution followed by motion towards the wall for the second half of the revolution. In this way we maintain a similar approach as the smooth cylinder rotation. We use the same velocity definitions as above. Starting with the Taylor expansion about $L = 0$ we have the same drag coefficient as before Eq. \ref{eq:Jdrag} and the lift coefficient as,
\begin{align}
    \hat{f}_z^{\omega} = \varepsilon^{-1}\left(4 L-6 L^2+\frac{107 L^3}{10} + \mathcal{O}(L^4)\right). 
\end{align}
Thus the leading order approximation is $\hat{f}_z^{\omega} \approx 4  \varepsilon^{-1} L$ and the drag is still $\hat{f}_z^{\bot} \approx -3\sqrt{2}\pi\varepsilon^{-3/2}$. Setting $\hat{f}_z^{\omega}\eta r \Omega = \hat{f}_z^{\bot} \eta V^{\bot}$ gives the displacement during the first half of the revolution,
\begin{equation}
    \Delta_1 \sim \delta_0^{-1/2}r^{1/2}\ell.
\end{equation}
The second half of the revolution will follow the same derivation with a new gap, $\delta_1$, which will be larger than $\delta_0$. The wall normal force will now be towards the wall. Thus the net migration will be the summation of both displacements,
\begin{align}
    \Delta_1+\Delta_2 = \delta_0^{-1/2}r^{1/2}\ell - \delta_1^{-1/2}r^{1/2}\ell.
\end{align}
Replacing $\delta_1 = \delta_0 +\Delta_1$ and some rearranging one obtains the wall normal displacement scaling,
\begin{align}    
    \Delta \sim \ell \sqrt{\frac{r}{\delta_0}} - \frac{\ell \sqrt{\frac{r}{\delta_0}}}{\sqrt{1 + \ell r^{1/2} \delta_0^{-3/2}}}.
\end{align}

\section{Scaling law correction for pattern wall} \label{sec:scale-correct}

\begin{figure}
    \centering
    \includegraphics[width=0.8\columnwidth]{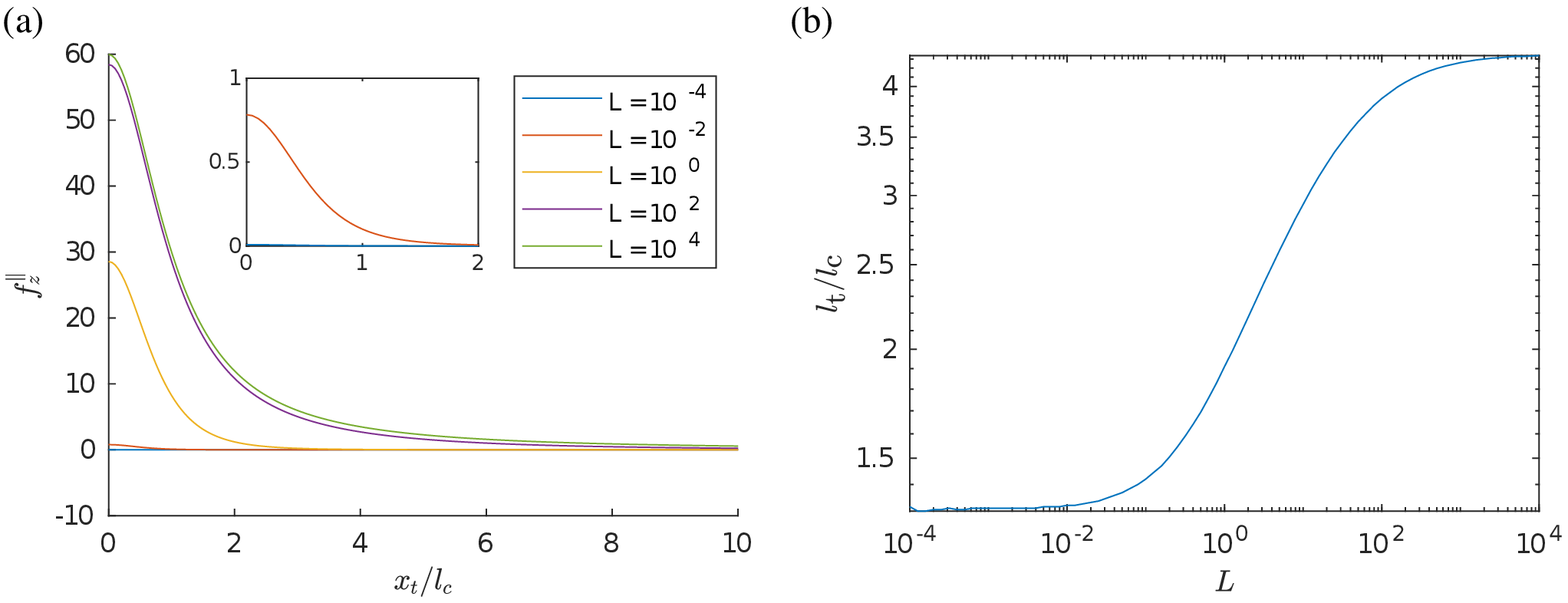}
    \caption{(a) The non-dimensional lift force as a function of distance from the surface transition point ($x_{\textrm{t}}$). For non-dimensional slip length $L<10^{-2}$ the standard definition for contact length ($l_{\textrm{c}}$) is a reasonable approximation (highlighted with the inset). For larger slip lengths the force is non-negligible with $x_{\textrm{t}}/l_{\textrm{c}}>1$. (b) Transition length as a function of non-dimensional slip length as defined in Eq. \ref{eq:trans_length}. }
    \label{fig:fz_xt}
\end{figure}

The simple scaling for the single transition proposed in the main paper,
$\Delta \sim \ell$, is insufficient to explain the net migration away from a patterned wall.
This scaling implies equal amount of upwards displacement for slip no-slip transition as
the preceding downwards displacement for no-slip slip transition.
%
%
In reality the migration could be a result of a more complex coupling between forces and effective time over which they act.
This effect could be captured by modifying the length scale over which the the
lift force approximately acts, see main paper Eq. 7. Here we call this length scale as ``transition length''. This would in turn
for the same translation velocity modify the effective time over which the lift
force is approximately acting on the cylinder.
This modification is motivated by analytic solution of the lift coefficient
configuration, where the cylinder center is moved away from the transition point.
In Fig. \ref{fig:fz_xt}(a) we show results for various dimensionless slip length
$L = \ell/\delta_0$ values. It is evident that as $L$ changes, the transition length also changes. If physical slip length $\ell$ is kept
constant, as for the patterned wall considered in the main paper, the $L$ value
decreases if gap thickness $\delta_0$ is increased. Consequently the transition length
shrinks when the cylinder is pushed away during the preceding slip no-slip transition, leading to smaller displacement towards the wall over next transition, even if the effective force is similar.

%
We define the transition length with a
help of the integral quantity,
\begin{equation}
l_{\textrm{t}} = \frac{l_{\textrm{c}}}{f_z^{\parallel,max}} \int_{-\infty}^{\infty} f_z^{\parallel}\,\textrm{d}X, \label{eq:trans_length}
\end{equation}
where $f_z^{\parallel,max}$ is the maximum value of the lift force obtained when
the cylinder is directly above the transition point. This integral measure provides
an estimate of an effective length over which a constant maximal force could be expected
to apply for capturing the displacement in the wall normal direction. Figure \ref{fig:fz_xt}(b) shows $l_{\textrm{t}}$ as a function of $L$ in a log-log plot. From this figure one can obtain a power law relationship, $l_{\textrm{t}} = l_{\textrm{c}} L^{\beta}$. Considering small $L$ one obtains $\beta \approx 0.003$ suggesting only a minor modification to the original scaling is required to capture the more complex patterned wall dynamics.    
%
%
%
With the new transition length the displacement estimate for a single slippage transition
is modified as follows.
As in the main paper, we estimate cylinder velocities as $V^{\parallel} \sim l_{\textrm{t}}/\tau$ and $V^{\bot} \sim \Delta/\tau$, where $\tau$ is the time scale. We take the leading order terms in Eqs. \ref{eq:fzp} and \ref{eq:drag} and setting the wall normal forces equal $f_z^{\parallel}\eta V^{\parallel} = f_z^{\bot}\eta V^{\bot}$. This gives the scale estimate for migration of a cylinder over one surface transition as, 
\begin{align}
      \Delta \sim \delta_0^{-0.003}\ell^{1.003}. \label{eq:pattern}
\end{align}
This scaling estimate predicts a smaller displacement for a larger gap thickness
$\delta_0$ and is consequently in agreement with net migration away from the wall.

\section{Numerical Method} \label{sec:num-meth}
\begin{figure}
    \centering
    \includegraphics[width=0.5\columnwidth]{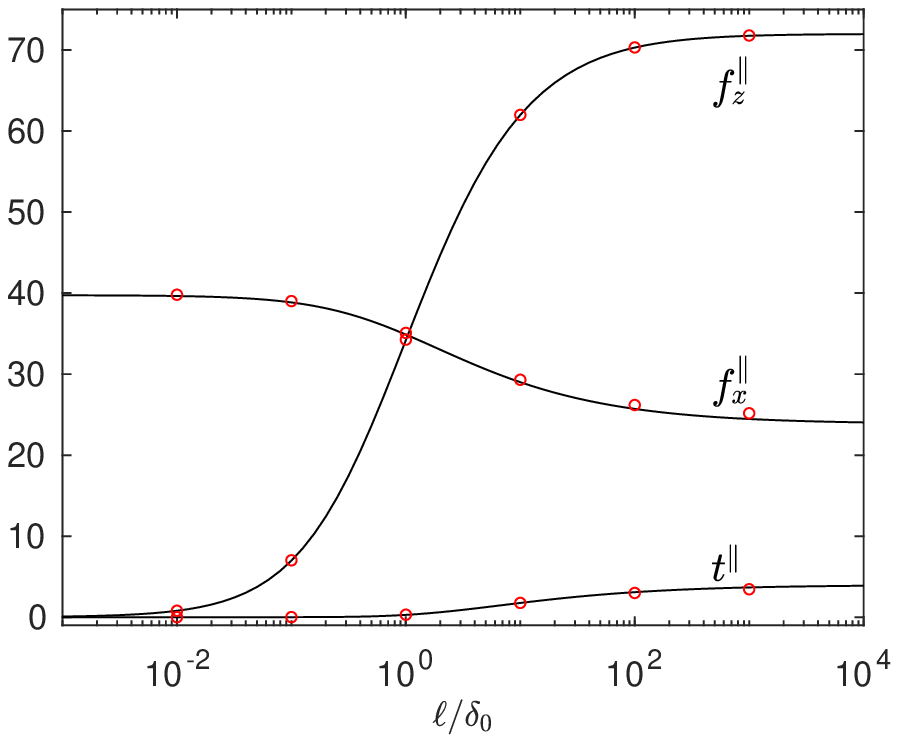}
    \caption{Comparison of the analytical force coefficients for wall parallel motion to FEM simulations. Here analytical solutions are shown in black and FEM simulations are shown with red circles.}
    \label{fig:res_coef_num}
\end{figure}

A finite element solver (FreeFem++) is used for the numerical component of this research \cite{freefem}. We use quadratic elements for the velocity components, $u, w$ and linear elements for pressure, $p$. Stokes equations are solved coupled with an Euler forward time stepping scheme. At each time step Newton's equations are solved for the cylinder velocity, $\bm{V}$ and rotation $\Omega$, from
\begin{equation}
    \rho_{\textrm{c}} V_{\textrm{c}} \frac{\textrm{d} \bm{V}}{\textrm{d} t} = \int_S{\bm{\sigma} \cdot \bm{n}\textrm{d}S} + V_c \left( \rho_{\textrm{c}} - \rho \right) \bm{g},\label{eq:newton}
\end{equation}
and
\begin{equation}
     r \rho_{\textrm{c}} I_{\textrm{c}} \frac{\textrm{d} \Omega}{\textrm{d} t} = \int_S{\bm{r}\times \left( \bm{\sigma} \cdot \bm{n}\right)\textrm{d}S} .\label{eq:newton2}
\end{equation}
The following  values are used in the numerical simulation; fluid density $\rho = \SI{970}{\kilo\gram\per\cubic\meter}$, fluid viscosity $\eta = \SI{1}{\pascal\second}$, particle density $\rho_c= \SI{8510}{\kilo\gram\per\cubic\meter}$, particle diameter $d = \SI{0.0127}{\meter}$, volume per length $V_c = \SI{1.13e-3}{\square\meter}$, second moment of inertia per length $I_c = \SI{1.73e-5}{\cubic \meter}$. The computational domain is re-meshed after each time step. The domain has stress-free boundary conditions for the fluid-fluid boundaries, Dirichlet conditions for the no-slip surfaces and Neumann conditions for the slip surfaces. The mesh is defined to maintain a minimum number of elements between the particle and wall to ensure good resolution of the thin gap. A mesh convergence test was performed increasing the minimum number of elements in the gap from 12 to 16 produced a relative change in pressure on the order of $10^{-5}$ suggesting a sufficiently converged solution. Finally, in Fig. \ref{fig:res_coef_num}, we compare the resistance coefficients obtained through FEM simulations with the analytical solutions, where we observe a good agreement ($2\%$ relative error), over a large range of slip lengths. The other resistance matrix components have comparable agreement.

\bibliography{CylinderMotionNearWall_SI}